\documentclass[twocolumn, secnumarabic, amssymb, amsmath,
nofootinbib,tightenlines, showpacs, nobibnotes, aps, prl]{revtex4}
\usepackage{amsfonts}

\usepackage{docs}%
\usepackage{bm}%
\usepackage[colorlinks=true,linkcolor=blue]{hyperref}%
\usepackage{graphicx}
\usepackage{dcolumn}

\begin{document}

\title{Supersolid in Bose-Bose-Fermi Mixtures subjected to a Square Lattice}
\author{Zhongbo Yan$^1$}
\author{Xiaosen Yang$^{2}$}
\author{Shaolong Wan$^1$}
\email[]{slwan@ustc.edu.cn} \affiliation{$^{1}$Institute for Theoretical
Physics and Department of Modern Physics University of Science and
Technology of China, Hefei, 230026, P. R. China\\
$^2$Beijing Computational Science Research Center, Beijing,
100084, P. R. China}
\date{\today}

\begin{abstract}
Two-component Bose condensates with repulsive interaction are
stable when $g_{\rm \scriptscriptstyle 1} g_{\rm
\scriptscriptstyle 2}<g_{\rm \scriptscriptstyle 12}^{2}$ is
satisfied. By tuning the interactions, we show that the
instability corresponding to bose-bose phase separation always
happens at a higher temperature than corresponding to bose-fermi
phase separation happens. Moreover, we find both the transition
temperature $T_{\rm \scriptscriptstyle DW}$ of supersolid and the
coherence peak at $k_{\rm \scriptscriptstyle DW}$ are enhanced in
the mixtures studied. These will make the observation of
supersolid in experiments more reachable.
\end{abstract}


\maketitle

\section{Introduction}

Supersolids, a concept simultaneously exhibiting superfluidity and
crystalline order, have been studied intensely over five decades
\cite{O. Penrose, A. F. Andreev, G. V. Chester, A. J. Leggett}.
Theoretically, people mainly focus on lattice models of
interacting bosons and fermions such as the Hubbard model and its
various generalizations and have obtained many important results
by numerical analysis \cite{G. G. Batrouni, I. Titvinidze}.
Experimentally, Kim and Chan recently reported they found
nonclassical rotational inertia which should be an direct evidence
of supersolid based on Leggett's suggestion in solid $^{4}$He
\cite{E. Kim1, E. Kim2}, however, it has also been pointed out
that this observation may not be due to supersolid but due to
other reasons, such as an increase in shear modulus of bulk solid
helium \cite{J. Day}, and triggered an intense debate \cite{M.
Boninsegni, D. Y. Kim}.

Besides the study of supersolids in condensed matter systems,
ultracold atoms in optical lattices \cite{D. Jaksch} have emerged
as a parallel platform with highly controllability to study
supersolids. Trapped Bose-Einstein condensates with dipole-dipole
interaction can produce a ``roton" minimum in the excitation
spectrum \cite{D. H. J. O'Dell, L. Santos,Shai Ronen}, and this led to the
prediction of supersolid upon softening of the roton excitation
energy \cite{D. Kirzhnits, G. Blatter}. Recently, on the basis of
off-resonant dressing of atomic Bose-Einstein condensates to
high-lying Rydberg states, people have found the effective atomic
interactions resulting from such a Rydberg dressing can also
produce a roton minimum and, therefore, provide a clean
realization of available model for supersolidity \cite{N. Henkel,
F. Cinti}.

In this work, we consider the two kind of bosons are two hyperfine
state of $^{87}$Rb, and the fermions are a hyperfine state of
$^{40}$K and investigate bose-bose-fermi mixtures in a square
lattice. For the bose-fermi mixtures subjected to a square
lattice, it has been pointed out that the density wave instability
introduced by fermions will establish crystalline order, while the
condensate bosons exhibit superfluidity, so a supersolid phase
emerges at finite temperature \cite{G. Blatter}. For the
bose-bose-fermi mixtures studied here, besides the density wave
instability introduced by fermions, there is another instability
between the two-component bose-condensates when $g_{\rm
\scriptscriptstyle 1}g_{\rm \scriptscriptstyle 2}=g_{\rm
\scriptscriptstyle 12}^{2}$, where $g_{\rm \scriptscriptstyle
1,2}$ are the repulsive intraspecies interaction and $g_{\rm
\scriptscriptstyle 12}$ is the interspecies interaction \cite{E.
Timmermans}. When $g_{\rm \scriptscriptstyle 1}g_{\rm
\scriptscriptstyle 2}>g_{\rm \scriptscriptstyle 12}^{2}$, the
bose-condensates are mixed and stable. When $g_{\rm
\scriptscriptstyle 1} g_{\rm \scriptscriptstyle 2}<g_{\rm
\scriptscriptstyle 12}^{2}$, the bose-condensates are unstable and
tend to either phase separation or collapse depending on $g_{\rm
\scriptscriptstyle 12} > 0$ or $< 0$. In this article, we assume
the bose-condensates are initially mixed and stable, and we find
that bose-bose phase separation always happens before bose-fermi
phase separation when we decrease the temperature. Moreover, we
find both the transition temperature $T_{\rm \scriptscriptstyle
DW}$ of supersolid and the coherence peak at $k_{\rm
\scriptscriptstyle DW}$ are enhanced comparing to the bose-fermi
mixtures case \cite{G. Blatter}.

The article is organized as follows. In Sec.\ref{sec2}, we
consider the two kind of bosons are two hyperfine state of
$^{87}$Rb, and the fermions are a hyperfine state of $^{40}$K and
investigate bose-bose-fermi mixtures in a square lattice, and give
the fermionic response in the static limit. In Sec.\ref{sec3},
we give the details of the instabilities and different phases induced
by the instabilities, and give a mean field description of the supersolid phase.
Some conclusions are obtained in Sec.\ref{sec4}.

\section{The bose-bose-fermi mixtures in a square lattice}
\label{sec2}

The Hamiltonian for the bose-bose-fermi mixtures takes the form
$H=H_{0}+H_{int}$ with ($\alpha=\uparrow,0,\downarrow$)

\begin{eqnarray}
H_{0}&=&\sum_{\alpha}\int d{\bf x}\psi_{\alpha}^{\dag}
\left\{(-\frac{\hbar^{2}}{2m_{\alpha}}
\bigtriangledown^{2} + V_{\alpha}(x) \right\} \psi_{\alpha}, \nonumber \\
H_{int}&=&\int d{\bf x} \left\{ g_{\rm \scriptscriptstyle
1}\psi_{\uparrow}^{\dag}\psi_{\uparrow}^{\dag}
\psi_{\uparrow}\psi_{\uparrow}+g_{\rm \scriptscriptstyle
2}\psi_{\downarrow}^{\dag}
\psi_{\downarrow}^{\dag}\psi_{\downarrow}\psi_{\downarrow} \right. \label{1} \\
& & \left. +2g_{\rm \scriptscriptstyle
12}\psi_{\uparrow}^{\dag}\psi_{\downarrow}^{\dag}\psi_{\downarrow}
\psi_{\uparrow}+2g_{\rm \scriptscriptstyle BF} \left(
\psi_{\uparrow}^{\dag}\psi_{\uparrow}+
\psi_{\downarrow}^{\dag}\psi_{\downarrow} \right)
\psi_{0}^{\dag}\psi_{0} \right\}, \nonumber
\end{eqnarray}
where $\psi^{\dag}_{\uparrow,\downarrow}$ are the bosonic field
operators and $\psi^{\dag}_{0}$ is the fermionic field operator.
In order to assure the mixtures to be stable, we assume all of the
interactions between bosons are repulsive, with $g_{\alpha \beta}
= 4 \pi a_{s, \alpha \beta} \hbar^{2}/m$ ($\alpha, \beta = 1, 2$.
In this work, we use $\alpha,\beta$ to label the interactions,
densities and phases of bosons, and use $\uparrow, \downarrow$
only to label the bosonic operators, moreover, when $\alpha=\beta$
we only keep $\alpha$ for convenience) and $g_{\rm
\scriptscriptstyle 1} g_{\rm \scriptscriptstyle 2}>g_{\rm
\scriptscriptstyle 12}^{2}$. $g_{\rm \scriptscriptstyle BF}=2\pi
a_{\rm \scriptscriptstyle BF} \hbar^{2}/\mu$, the strength of
coupling between bosons and fermions, we assume that they are
equal for both component of bosons. $a_{s,\alpha}$ is the
intraspecies scattering length, $a_{s,12}$ is the interspecies
scattering length, and $\mu$ is the relative mass.
$V_{\alpha}({\bf x})=V_{\alpha}[\sin^{2}(\pi x/a)+\sin^{2} (\pi
y/a)]$ is the periodic potential produced by the optical lattice
with wave-length $\lambda=2a$ and $m_{\alpha}$ are the mass of
bosons and fermions. As the bosons are two hyperfine state of
$^{87}$Rb, it is justified to assume $m_{\uparrow}=m_{\downarrow}$
and $V_{\uparrow}=V_{\downarrow}$ for simplicity in the following.
Since the fermions are single component, the interaction between
them can be neglected due to Pauli exclusion principle.

In order to obtain the Hamiltonian in momentum space, we follow
the procedures used in Ref.\cite{G. Blatter} and expand the
bosonic and fermionic field operators $\psi_{\alpha}$ in the forms
\begin{eqnarray}
\psi_{\uparrow,\downarrow}(x) &=&\sum_{{\bf k}\in K}b_{{\bf
k}\uparrow,\downarrow}w_{{\bf k} \uparrow,\downarrow}(x),
\nonumber\\
\psi_{0}(x)&=&\sum_{{\bf k}\in K}c_{{\bf k}}v_{{\bf k}}(x),
\label{2}
\end{eqnarray}
where $K$ denotes the first Brillouin zone, $b_{{\bf
k}\uparrow,\downarrow}$ and $c_{{\bf k}}$ are the bosonic and
fermionic annihilation operators, while $w_{{\bf
k}\uparrow,\downarrow}(x)$ and $v_{{\bf k}}(x)$ are the Bloch wave
functions corresponding to a single boson ($\uparrow$ or
$\downarrow$) or fermion in the periodic potential $V_{\alpha}$,
respectively. Since $m_{\uparrow} =m_{\downarrow}$ and
$V_{\uparrow}=V_{\downarrow}$,  $w_{{\bf k}\uparrow}(x)$ should be
equal to $w_{{\bf k}\downarrow}(x)$. Therefore, we use $w_{{\bf
k}}(x)$ to denote both of them for convenience. Substituting
Eq.(\ref{2}) into Eq.(\ref{1}) and restricting in the lowest Bloch
band, we obtain the Hamiltonian in momentum space as
\begin{eqnarray}
H&=&\sum_{{\bf k}\in K,\sigma}\epsilon_{\rm \scriptscriptstyle  B\sigma}b_{{\bf k}\sigma}^{\dag}
b_{{\bf k}\sigma}+ \sum_{\{{\bf k,k',q,q',\sigma}\}}\frac{U_{\scriptscriptstyle \rm B\sigma}}{2N}
b^{+}_{{\bf k}\sigma}b^{}_{{\bf k}'\sigma}b^{+}_{{\bf q}\sigma} b^{}_{{\bf q}'\sigma} \nonumber\\
& &+\sum_{\{{\bf k,k',q,q'}\}}\frac{U_{\scriptscriptstyle \rm B,12}}{N}b^{+}_{{\bf k}\uparrow}b^{}
_{{\bf k}'\uparrow}b^{+}_{{\bf q}\downarrow} b^{}_{{\bf q}'\downarrow}+\sum_{{\bf q} \in K} \epsilon_
{\scriptscriptstyle \rm F}({\bf q}) c^{+}_{{\bf q}}c^{}_{{\bf q}} \nonumber \\
& &+\frac{U_{\scriptscriptstyle \rm BF}}{N}\sum_{\{{\bf k,k',q,q',\sigma}\}}b^{+}_{ {\bf k}\sigma}
b^{}_{{\bf k}'\sigma} c^{+}_{{\bf q}}c^{}_{{\bf q}'}, \label{3}
\end{eqnarray}
where $N$ is the number of unit cells,
$\epsilon_{\scriptscriptstyle \rm F,B\sigma}({\bf k})$ denote the
energy dispersion of the fermions and bosons, respectively, while
$U_{\scriptscriptstyle \rm BF}= g_{ \scriptscriptstyle \rm BF}\int
d{\bf x}|\widetilde{w}|^{2}|\widetilde{v}|^{2}$ and
$U_{\scriptscriptstyle \rm B,\alpha\beta}= g_{\scriptscriptstyle
\rm \alpha\beta}\int d{\bf x}|\widetilde{w}|^{4}$, with
$\widetilde{w} ({\bf x})$ and $\widetilde{v}({\bf x})$, the
Wannier functions associated with the Bloch band $w_{{\bf k}}({\bf
x})$ and $v_{{\bf k}}({\bf x})$. In a deep optical lattice, the
Wannier functions $\widetilde{w}({\bf x})$ and $\widetilde{v}({\bf
x})$ are well localized around the minimum of $V_{\alpha}$. As a
result, the Hamiltonian reduces to a familiar Bose-Fermi-Hubbard
model, and for $\epsilon_{\scriptscriptstyle \rm F,B\sigma}({\bf
k})$, only nearest neighbor hopping survives,
\begin{eqnarray}
 \epsilon_{\scriptscriptstyle
      \rm B}({\bf q}) &=& 2 J_{\rm \scriptscriptstyle B} \left[2- \cos
      \left(q_{x} a\right)- \cos \left( q_{y} a \right)\right], \nonumber \\
      \epsilon_{\scriptscriptstyle \rm F}({\bf q}) &=& -2 J_{\rm
    \scriptscriptstyle F} \left[ \cos \left(q_{x} a\right)
     +    \cos \left( q_{y} a\right) \right],
         \label{4}
\end{eqnarray}
where $J_{\rm \scriptscriptstyle B,F}$ is the hopping energy for
fermions and bosons, respectively. The bosonic dispersion relation
implies $\mu_{B}=-4J_{B}$ and the bosons will form a zero-momentum
Bose-Einstein condensation for sufficiently low temperature. The
fermionic dispersion relation implies the Fermi surface at
half-filling $n_{\rm \scriptscriptstyle F}=1/2$ (where $\mu_{\rm
F}=0$, in this work. $n_{\rm \scriptscriptstyle F}$ and $n_{\rm
\scriptscriptstyle B}$ denote the number of particles per unit
cell) and exhibits perfect nesting for ${\bf k}_{\rm
\scriptscriptstyle DW}=(\pi/a,\pi/a)$ and van Hove singularities
at ${\bf k}=(0,\pm \pi/a),(\pm\pi/a,0)$.

Integrating out the fermions produces two effects. To the first
order in $U_{\scriptscriptstyle \rm BF}$ (in this work, we focus
on weak interaction, so an expansion in $U_{\scriptscriptstyle \rm
BF}$ and a cut at the second order are justified), the fermions
simply produce a (trivial) shift of the bosonic chemical potential
$\mu_{\scriptscriptstyle \rm
B\sigma}\rightarrow\mu_{\scriptscriptstyle \rm B\sigma}-U_{
\scriptscriptstyle \rm BF}n_{\scriptscriptstyle \rm F}$. To the
second order in $U_{\scriptscriptstyle \rm BF}$, the fermions
provide an effective interaction for the bosons which depends on
the temperature $T$ of the fermionic atom gas,
\begin{eqnarray}
    H_{\rm \scriptscriptstyle int} &=&\frac{1}{2N}\sum_{\{{\bf k,k',q,q',\sigma}\}}
    U_{\rm \scriptscriptstyle B\sigma}(T,{\bf q-q'})b^{+}_{{\bf k}\sigma} b^{}_{{\bf k}'\sigma}
    b^{+}_{{\bf q}\sigma}b^{}_{{\bf q}'\sigma} \nonumber\\
    & &+\frac{1}{N}\sum_{\{{\bf k,k',q,q',}\}}U_{\rm \scriptscriptstyle B,12}(T,{\bf q-q'})
    b^{+}_{{\bf k}\uparrow} b^{}_{{\bf k}'\uparrow} b^{+}_{{\bf q}\downarrow}b^{}_{{\bf q}'\downarrow}, \nonumber
\end{eqnarray}
with
\begin{eqnarray}
U_{\rm \scriptscriptstyle B\sigma}(T,{\bf q-q'})&=&U_{\rm \scriptscriptstyle B\sigma}
    + U_{\rm \scriptscriptstyle FB}^{2} \chi(T,{\bf q-q'}), \nonumber \\
U_{\rm \scriptscriptstyle B,12}(T,{\bf q-q'})&=&U_{\rm \scriptscriptstyle B,12}
    + U_{\rm \scriptscriptstyle FB}^{2} \chi(T,{\bf q-q'}).\label{5}
\end{eqnarray}
The fermionic response in the static limit is given by the
Lindhard function
\begin{equation}
    \chi(T,{\bf q}) =  \int_{K}\frac{d{\bf k}}{v_{0}}
    \frac{f[\epsilon_{\rm \scriptscriptstyle F}({\bf k})]
    -f[\epsilon_{\rm \scriptscriptstyle F}({\bf k+q})] }
    {\epsilon_{\rm \scriptscriptstyle F}({\bf k})-
    \epsilon_{\rm \scriptscriptstyle F}({\bf k+q}) + i \eta},
    \label{6}
\end{equation}
where $v_{0}=( 2\pi/a)^{2}$ is the volume of the first Brillouin
zone, $f(\epsilon)=1/[1+\exp(\epsilon/T)]$ ($\mu_{\rm
\scriptscriptstyle F}=0$ at half filling) is just the Dirac-Fermi
distribution function. The static limit is justified if the
fermions are much faster than the bosons ($J_{\rm
\scriptscriptstyle F}>>J_{\rm \scriptscriptstyle B}$), as then the
fermionic response occurs on much faster timescales than the
movement of the bosons, and one can safely neglect retardation
effects \cite{Peter P. Orth}. Using the fermionic dispersion
relation Eq.(\ref{4}), the Lindhard function exhibits two
logarithmic singularities at ${\bf q}=0$ and ${\bf k}_{\rm
\scriptscriptstyle DW}$. The singularity at ${\bf q}=0$ is purely
due to the logarithmic van Hove singularity in the density of
states, and the singularity at ${\bf k}_{\rm \scriptscriptstyle
DW}$ is due to the combination of van Hove singularities and
perfect nesting. The singularity at ${\bf q}=0$ induces an
instability towards a series of phase separation, while the
singularity at ${\bf k}_{\rm \scriptscriptstyle DW}$ induces an
instability towards density wave formation and provides a
supersolid phase. The two instabilities are competing with each
other.

\section{INSTABILITIES AND PHASES}
\label{sec3}

For the weak interaction, when the temperatures is well below the
superfluid transition temperature $T_{\rm \scriptscriptstyle KT}$
of the bosons, the Lindhard function at ${\bf q}=0$  reduces to
$\chi(T \rightarrow 0,0)$ and takes the form \cite{G. Blatter}
\begin{equation}
    \chi(T\rightarrow 0,0)= \int d\epsilon N(\epsilon)
     \partial_{\epsilon} f(\epsilon)
    \sim-N_{0}
    \ln\frac{16 c_{1} J_{\rm \scriptscriptstyle F}}{T}, \label{7}
\end{equation}
with $N(\epsilon)\sim N_{0} \ln|16 J_{\rm \scriptscriptstyle F}
/\epsilon | $, $N_{0}= 1/(2\pi^{2} J_{\rm \scriptscriptstyle F})$
and $c_{1} =2\exp(C)/\pi\approx 1.13$. As $\chi(T\rightarrow 0,0)$
is always negative, the coupling between the bosons and the
fermions induces an attractive interaction, which is proportional
to $U_{\rm \scriptscriptstyle FB}^{2} \chi(T,{\bf 0})$, between
the bosons (see Eq.(\ref{5})). This attractive interaction has the
effect to reduce the repulsive interactions $U_{B,\alpha\beta}$
between bosons to $U_{\rm \scriptscriptstyle eff,\alpha\beta}=
U_{\rm \scriptscriptstyle B,\alpha\beta} + U_{\rm
\scriptscriptstyle FB}^{2} \chi(T,0)$. As a result, even $U_{\rm
\scriptscriptstyle B,1}U_{\rm \scriptscriptstyle B,2}>U_{\rm
\scriptscriptstyle B,12}^{2}$ (equivalent to $g_{\rm
\scriptscriptstyle 1} g_{\rm \scriptscriptstyle 2}>g_{\rm
\scriptscriptstyle 12}^{2}$) initially, $U_{\rm \scriptscriptstyle
eff,1}U_{\rm \scriptscriptstyle eff,2}$ can be tuned to equal to
$U_{\rm \scriptscriptstyle eff,12}^{2}$ by lowering the
temperature to some value. Moreover, a superfluid condensate at
low temperatures to be stable requires a positive effective
interaction $U_{\rm\scriptscriptstyle eff,\alpha}>0$. If we take
$U_{\rm \scriptscriptstyle B,1}$ as the energy unit, and define
the ratios $U_{\rm \scriptscriptstyle B,2}/U_{\rm
\scriptscriptstyle B,1}$, $U_{\rm \scriptscriptstyle B,12}/ U_{\rm
\scriptscriptstyle B,1}$ and $U_{\rm \scriptscriptstyle BF}
/U_{\rm \scriptscriptstyle B,1}$ as $\gamma$, $\lambda$ and
$\kappa$, respectively. The condition
$U_{eff,12}^{2}=U_{eff,1}U_{eff,2}$ defines the critical
temperature $T_{\rm \scriptscriptstyle BB,PS}$ for bose-bose phase
separation,
\begin{equation}
    T_{\rm \scriptscriptstyle BB,PS} = 16 c_{1}
        J_{\rm \scriptscriptstyle F} \exp\left[
        \frac{\lambda^{2}-\gamma}{N_{0}\kappa^{2}(1+\gamma-2\lambda)} \right]. \label{8}
\end{equation}
The condition $U_{\rm \scriptscriptstyle eff,\alpha}=0$ defines
two critical temperatures $T_{\rm \scriptscriptstyle BF1,PS}$ and
$T_{\rm \scriptscriptstyle BF2,PS}$ for bose-fermi phase
separation.
\begin{eqnarray}
    T_{\rm \scriptscriptstyle BF1,PS} &=& 16 c_{1}
        J_{\rm \scriptscriptstyle F} \exp\left[-
        \frac{1}{N_{0}\kappa^{2}} \right], \nonumber \\
    T_{\rm \scriptscriptstyle BF2,PS} &=& 16 c_{1}
        J_{\rm \scriptscriptstyle F} \exp\left[-
        \frac{\gamma}{N_{0}\kappa^{2}} \right].   \label{9}
\end{eqnarray}

\begin{figure}
\includegraphics[width=8.5cm, height=6.0cm]{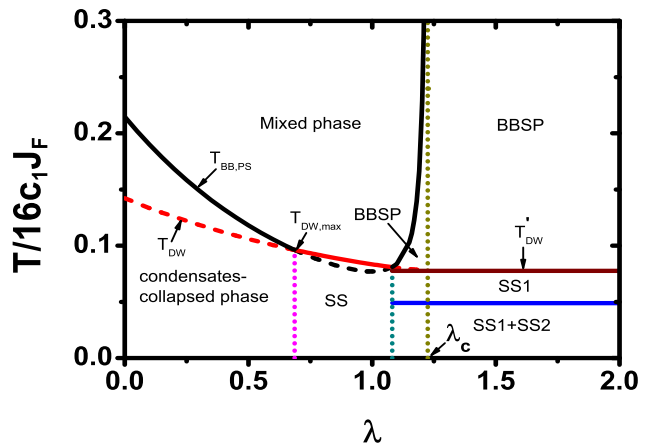}
\caption{(Color online) $\lambda - T$ Phase diagram. Parameters
are set as $t_{\rm B} = 0.55$, $N_{0} \kappa^{2} = 0.39$, $\gamma
= 1.5$, $\lambda_{c} = \sqrt{\gamma}$. For $\lambda >
\lambda_{c}$, the two-component bosons are phase separation
initially (BBPS), and supersolid corresponding to
bosons-$\uparrow$ (SS1) emerges when the temperature is below
$T'_{DW}$. For $\lambda < \lambda_{c}$, the supersolid phase
established by the two Bose-condensates appears when the
temperature is below $T_{\rm DW}$ with $T_{\rm BB, PS} < T_{\rm
DW}$. For $\lambda < 1$ ($1< \lambda < \lambda_{c}$), the two
Bose-condensates collapse (separate) when the temperature is below
$T_{\rm BB, PS}$ with $T_{\rm BB, PS} > T_{\rm DW}$.}
\label{Fig.1}
\end{figure}

When $\lambda\neq1$ and $\gamma$, it is directly to show that
$(\gamma-\lambda^{2})/(1+\gamma-2\lambda)$ is always smaller than
$\min\{1,\gamma\}$ under the constraint $\lambda^{2}<\gamma$,
which is the condition that the bose condensates are initially
mixed (see Fig.\ref{Fig.1}). $(\gamma-\lambda^{2})/(1+\gamma-2
\lambda)<\min\{1,\gamma\}$ indicates when we lower the
temperature, the bose-bose mixtures are always easier to be
unstable and phase separated (or collapse, see Fig.\ref{Fig.1})
than the bose-fermi mixtures. Moreover, when $\lambda$ gets close
to the boundary $\sqrt{\gamma}$,
$(\gamma-\lambda^{2})/(1+\gamma-2\lambda)$ decreases very fast, as
a result, $T_{\rm \scriptscriptstyle BB,PS}$ increases
exponentially to values much larger than $max\{T_{\rm
\scriptscriptstyle BF1,PS}, T_{\rm \scriptscriptstyle BF2,PS}\}$
and easy to reach in experiments. Therefore, such a bose-bose
phase separation induced by fermions should  be easy to be
observed in experiments. If we continue to lower the temperature
after the bose-bose mixtures are phase separated, we can expect
that bose-fermi phase separation will happen and all the
components will distribute separately in space at last.

Now, we discuss the second instability induced by the singularity
in the Lindhard function at ${\bf k}_{\rm \scriptscriptstyle DW}$.
Using Eq.(\ref{6}) and the perfect nesting
$\epsilon_{\scriptscriptstyle \rm F} ({\bf q}+{\bf k}_{\rm
\scriptscriptstyle DW})\!= - \epsilon_{\scriptscriptstyle \rm
F}({\bf q})$, the Lindhard function becomes \cite{G. Blatter}
\begin{eqnarray}
\chi(T,{\bf k}_{\rm DW}) = \int d \epsilon N(\epsilon)
     \frac{\tanh\left({\epsilon}/{2T}\right)}
     {-2 \epsilon} \sim -\frac{N_{0}}{2}
   \left[\ln \frac{16 c_{1} J_{\rm \scriptscriptstyle F}}{T}
   \right]^{2}. \nonumber \\
\end{eqnarray}
The combination of van Hove singularities and
perfect nesting produces a $[\ln T]^{2}$ singular behavior. Such a
singular behavior can produce a roton minimum at ${\bf k}_{\rm
\scriptscriptstyle DW}$. Within Bogoliubov theory, the bosonic
quasi-particle spectrum becomes
\begin{eqnarray}
    E_{\rm \scriptscriptstyle B,\pm}^{2}({\bf q}) &=& \epsilon_{\rm \scriptscriptstyle B}^{2}
    ({\bf q})+\epsilon_{\rm \scriptscriptstyle B}({\bf q})n_{\rm \scriptscriptstyle B}
    \left[U_{\rm \scriptscriptstyle B1}(T,q) + U_{\rm \scriptscriptstyle B2}(T,q)\right] \nonumber \\
    &&\pm \left\{\epsilon_{\rm \scriptscriptstyle B}^{2}({\bf q} )n_{\rm \scriptscriptstyle B}^{2}
    \left[U_{\rm \scriptscriptstyle B1}(T,q) - U_{\rm \scriptscriptstyle B2}(T,q) \right]^{2} \right. \nonumber \\
    &&+ \left. 4\epsilon_{\rm \scriptscriptstyle B}^{2}({\bf q})n_{\rm \scriptscriptstyle B}^{2}
    U_{\rm \scriptscriptstyle B,12}^{2}(T,q) \right\}^{1/2}, \label{10}
\end{eqnarray}
here we have assumed $n_{\rm \scriptscriptstyle B1}=n_{\rm
\scriptscriptstyle B2} =n_{\rm \scriptscriptstyle B}$. The induced
attraction proportional to $U_{\rm \scriptscriptstyle
BF}^{2}\chi(T,{\bf k}_{\rm \scriptscriptstyle DW})$ reduces the
energy of quasi-particles at ${\bf k}_{\rm \scriptscriptstyle DW}$
from a pure-bosonic maximum (when $U_{\rm \scriptscriptstyle
BF}=0$, the maximum of $E_{\rm \scriptscriptstyle B,-}({\bf q})$
locates at ${\bf k}_{\rm \scriptscriptstyle DW}$) to an induced
zero roton minimum ($E_{\rm \scriptscriptstyle B,-}({\bf k}_{\rm
\scriptscriptstyle DW})) =0$) at the critical temperature
\begin{equation}
    T_{\rm \scriptscriptstyle DW} = 16 c_{1} J_{\rm \scriptscriptstyle F}
    \exp\left[ - \sqrt{\frac{t_{\rm \scriptscriptstyle B}^{2}+2t_{\rm
    \scriptscriptstyle B}(1+\gamma)+4\gamma-4\lambda^{2}}{2N_{0}
    \kappa^{2}(1+\gamma+t_{\rm \scriptscriptstyle
    B}-2\lambda)}}\right]
    \label{11}
\end{equation}
with $t_{\rm \scriptscriptstyle B}= 8J_{\rm \scriptscriptstyle B}
/n_{\rm \scriptscriptstyle B}U_{\rm \scriptscriptstyle B1}$. As
$E_{\rm \scriptscriptstyle B,-}({\bf k}_{\rm \scriptscriptstyle
DW}, T_{\rm \scriptscriptstyle DW})=E_{\rm \scriptscriptstyle
B,-}({\bf k=0})=0$, we can expect the boson modes $b_{{\bf k}_{\rm
\scriptscriptstyle DW} \alpha}$ to become macroscopically occupied
just like the boson mode $b_{0\alpha}$ below this critical
temperature. Comparing this result to the one obtained in
Ref.\cite{G. Blatter},
\begin{equation}
    T_{\rm \scriptscriptstyle DW}^{'} = 16 c_{1}J_{\rm \scriptscriptstyle F}
\exp\left[ - \sqrt{(2+t_{\rm \scriptscriptstyle B})/\lambda_{\rm
\scriptscriptstyle FB}}\right], \nonumber
\end{equation}
we find $T_{\rm \scriptscriptstyle DW}$ is always higher than
$T_{\rm \scriptscriptstyle DW}^{'}$ when parameters appearing in
both systems take the same values (see Fig.\ref{Fig.1}). Moreover,
since $T_{\rm \scriptscriptstyle DW}$ depends on
$\sqrt{\frac{t_{\rm \scriptscriptstyle B}^{2}+2t_{\rm
\scriptscriptstyle B}(1+\gamma)+4
\gamma-4\lambda^{2}}{2N_{0}\kappa^{2}(1+\gamma+t_{\rm
\scriptscriptstyle B}-2\lambda)}}$ exponentially (\ref{11}), a
small change of this term may induce a great change of $T_{\rm
\scriptscriptstyle DW}$. Therefore, such an enhancement of
critical temperature can be large. However, $T_{\rm
\scriptscriptstyle DW}$ can not increase as greatly as $T_{\rm
\scriptscriptstyle BB,PS}$, since $t_{\rm \scriptscriptstyle B}$
has to be larger than a critical value $t_{\rm \scriptscriptstyle
SF-MI}\approx 1/3$, below which Mott insulating phase emerges and
the above picture fails \cite{D. Jaksch}. As a comparison, we
calculate $T_{\rm \scriptscriptstyle DW}$ based on the parameters
used in Ref.\cite{G. Blatter} and find $T_{\rm \scriptscriptstyle
DW,max}/T_{\rm \scriptscriptstyle DW}^{'}\approx1.3$ (this ratio
goes to the maximum when $\gamma\rightarrow1$, in Fig.\ref{Fig.1},
we take $\gamma=1.5$ just for manifesting every phase) under the
constraint $T_{\rm \scriptscriptstyle DW,max}> T_{\rm
\scriptscriptstyle BB,PS}$ (if bose-bose phase separation happens
first, $T_{\rm \scriptscriptstyle DW}$ reduces to $T_{\rm
\scriptscriptstyle DW}^{'}$, and the enhancement effect of $T_{\rm
\scriptscriptstyle DW}$ misses). Such an enhancement of  $T_{\rm
\scriptscriptstyle DW}$ is of realistic meaning, since the lower
the temperature is, the harder it is to reach in cold atomic
experiments.

For temperatures well below $T_{\rm \scriptscriptstyle KT}$ and
$T_{\rm \scriptscriptstyle DW}$, both the boson mode $b_{{\bf k}
_{\rm \scriptscriptstyle DW}\alpha}$ and $b_{0\alpha}$ are
macroscopically occupied, therefore, it is justified to use mean
fields $\langle b_{{\bf k}_{\rm \scriptscriptstyle
DW}\alpha}\rangle$ and $\langle b_{0\alpha} \rangle$ to substitute
them. Introducing the mean fields $\langle b_{0\alpha}
\rangle=\sqrt{n_{0\alpha}N} \exp(i \varphi_{0\alpha})$ and
$\langle b_{{\bf k}_{\rm \scriptscriptstyle DW}\alpha}\rangle =
(\Delta_{\alpha} /2 U_{\rm \scriptscriptstyle BF})
\sqrt{N/n_{0\alpha}}\exp(i \varphi_{\alpha})$ with the constraint
$n_{\rm B\alpha}= n_{0\alpha} + \Delta_{\alpha}^{2}/(4n_{0\alpha}
U^{2}_{\rm \scriptscriptstyle BF})$ and neglecting thermal
excitations of bosonic quasi-particles \cite{G. Blatter}, we
obtain the bosonic densities as
\begin{equation}
        n_{\rm \scriptscriptstyle B\alpha}(x,y)=n_{\rm \scriptscriptstyle
        B\alpha}+ \frac{ \Delta_{\alpha} \cos \theta_{\alpha}}{ U_{\rm \scriptscriptstyle BF}}
        \left[\cos \frac{\pi x}{a} \cos \frac{\pi y}{a}\right]
        \label{densitywave}
\end{equation}
with $\theta_{\alpha}= \varphi_{0\alpha} - \varphi_{\alpha}$. The
phase difference $\Delta\theta=\theta_{1}-\theta_{2}$ between the
two bosonic density waves determines whether they are constructive
or destructive. Introducing $\langle b_{0\alpha} \rangle$ and
$\langle b_{{\bf k}_{\rm \scriptscriptstyle DW}\alpha}\rangle$ to
Hamiltonian (\ref{3}) and neglecting terms independent of
$\Delta_{\alpha}$, the Hamiltonian per unit cell is given as
\begin{eqnarray}
    \frac{H}{N}&=& 2 J_{\rm \scriptscriptstyle B} \frac{\Delta_{1}^{2}+\Delta_{2}^{2}}
    { n_{\rm \scriptscriptstyle B}U_{\rm \scriptscriptstyle BF}^{2}}
     +\frac{ U_{\rm \scriptscriptstyle B1}\Delta_{1}^{2} \cos^{2} \theta_{1}}
     {2 U_{\rm \scriptscriptstyle BF}^{2}}+\frac{ U_{\rm \scriptscriptstyle B2}
     \Delta_{2}^{2} \cos^{2} \theta_{2}}{2 U_{\rm \scriptscriptstyle BF}^{2}} \nonumber \\
     & &+\frac{ U_{\rm \scriptscriptstyle B,12} \Delta_{1}\Delta_{2}( \cos^{2} \theta_{1}+
     \cos^{2} \theta_{2}+2\cos(\theta_{1}- \theta_{2}))}{4 U_{\rm \scriptscriptstyle BF}^{2}} \nonumber \\
     & &+ \frac{H_{\rm \scriptscriptstyle F}}{N} + o(\Delta^{4}). \label{12}
\end{eqnarray}
The terms in the first and second lines describe the increase in
the kinetic and interaction energies of the bosons due to the
modulation of densities triggered by the boson modes $b_{{\bf
k}_{\rm \scriptscriptstyle DW}\alpha}$, while  $H_{\rm F}$ takes
the form
\begin{equation}
    H_{\rm F} =\frac{1}{2}\sum_{{\bf q} \in K}
    \left(\begin{array}{cc} c_{{\bf q}}^{+},&
    c_{{\bf q}'}^{+}\end{array}\right)\!
    \left( \begin{array}{cc} \epsilon_{\rm \scriptscriptstyle F}({\bf q}) &  \Delta(\theta_{1},\theta_{2}) \\
    \Delta(\theta_{1},\theta_{2}) &  \epsilon_{\rm \scriptscriptstyle F}({\bf q}')
    \end{array}\right)
    \left( \begin{array}{c} c^{}_{{\bf q}}\\ c^{}_{{\bf q}' }
    \end{array}\right)
\end{equation}
with a constraint ${\bf q}'= {\bf q} - {\bf k}_{\rm
\scriptscriptstyle DW}+{\bf K}_{h}$ (the reciprocal lattice vector
${\bf K}_{h}$ ensures the constraint ${\bf q}'\in K$) and
$\Delta(\theta_{1},\theta_{2})=\Delta_{1}
\cos\theta_{1}+\Delta_{2} \cos\theta_{2}$. Diagonalizing the
fermionic Hamiltonian, we obtain the fermionic quasi-particle
excitation spectrum $E_{\rm \scriptscriptstyle F}({\bf k},\Delta)=
\pm [\epsilon_{\rm \scriptscriptstyle F}^{2}({\bf k})+(\Delta_{1}
\cos\theta_{1}+\Delta_{2} \cos\theta_{2})^{2}]^{1/2}$. To
determine the phase difference $\Delta\theta$, we minimize the
thermodynamic potential $\Omega(T,\Delta_{1},\Delta_{2},
\theta_{1},\theta_{2})$ and find a constraint between $\theta_{1}$
and $\theta_{2}$: $\theta_{1} = \theta_{2} = s \pi$, with $s$ an
integer. Therefore, the phase difference $\Delta\theta=0$, the two
bosonic density waves are completely constructive and produce a
stronger density wave. A stronger density wave makes the
crystalline order favorable, therefore, such a phase-locking
effect is favorable to form a supersolid phase. As $\theta_{1} =
\theta_{2} = s \pi$, $\Delta_{+}=\Delta_{1}+\Delta_{2}$ is in fact
the gap. Introducing  $\Delta_{\pm}=\Delta_{1}\pm\Delta_{2}$ and
rewriting Eq.(\ref{12}), the self-consistency relations
($\partial_{\Delta_{\pm}} \Omega=0$) take the form
\begin{eqnarray}
& &(1+\gamma+t_{B}-2\lambda)\Delta_{-}=(\gamma-1)\Delta_{+},\nonumber \\
& &\frac{1}{2N_{0}\kappa^{2}}[t_{B}+(1+\frac{\Delta_{-}}{\Delta_{+}})+
\gamma(1-\frac{\Delta_{-}}{\Delta_{+}})+2\lambda]=\nonumber \\
& &\frac{1}{N_{0}}\int_{K}\frac{d{\bf k}}{v_{0}} \frac{\tanh
\left[E_{\rm \scriptscriptstyle F}({\bf k},\Delta_{+})/2
T\right]}{E_{\rm \scriptscriptstyle F}({\bf k},\Delta_{+})}.
\end{eqnarray}
Setting $\Delta_{+}(T_{\rm \scriptscriptstyle DW}) = 0$ and
combining the two equations above, we reproduce the critical
{temperature} in Eq.(\ref{11}). This confirms the picture that
upon softening $E_{\rm \scriptscriptstyle B,-}({\bf k}_{\rm
\scriptscriptstyle DW})$ to zero the bosonic density waves
characterized by $\langle b_{{\bf k}_{\rm \scriptscriptstyle
DW}\alpha}\rangle\neq0$ emerge with a breaking of the discrete
symmetry of the optical lattice be right. Furthermore, using the
density of states $N_{\Delta_{+}}(\epsilon)=N(\sqrt{\epsilon^2
-\Delta_{+}^2})|\epsilon|/\sqrt{\epsilon^2+\Delta_{+}^{2}}$, the
gap at $T=0$ becomes
\begin{equation}
    \Delta_{+}(0)= 32 J_{\rm \scriptscriptstyle F}\exp\left[  - \sqrt
    {\frac{t_{\rm B}^{2}+2t_{\rm B}(1+\gamma)+4\gamma-4\lambda^{2}}
    {2N_{0}\kappa^{2}(1+\gamma+t_{\rm B}-2\lambda)}} \right],
\end{equation}
and the standard BCS relation $2 \Delta_{+}(0)/ T_{\rm
\scriptscriptstyle DW}= 2 \pi/e^C\approx 3.58$ holds. This
relation implies that the density wave have the characteristic of
the superfluid, an evidence of supersolid. Therefore,
$T_{\rm\scriptscriptstyle DW}$ is just the critical temperature of
supersolid to emerge.

In experiments, the supersolid  can be detected via the usual
coherence peak of a bosonic condensate in an optical lattice. The
appearance of a coherence peak at ${\bf k}_{\rm \scriptscriptstyle
DW}$ is a symbol that the supersolid appears. Since the weight of
this coherence peak is proportional to the number of bosons
condensed at ${\bf k}_{\rm \scriptscriptstyle DW}$, the larger
$\Delta_{+}$ (here equivalent to $\langle b_{{\bf k}_{\rm
\scriptscriptstyle DW}\alpha}\rangle$) is, the sharper the peak
is. Therefore, based on the similarity of the forms between
$\Delta_{+}(0)$ and $T_{\rm\scriptscriptstyle DW}$, we find a
sharper coherence peak at ${\bf k}_{\rm \scriptscriptstyle DW}$
appears in bose-bose-fermi mixtures compared to the one appearing
in bose-fermi mixtures \cite{G. Blatter} when parameters appearing
in both systems take the same values. Based on the results above,
we can make the conclusion that it is more favorable to observe
the supersolid in bose-bose-fermi mixtures than in bose-fermi
mixtures.

\section{Conclusions}
\label{sec4}

In this article, we have investigated a bose-bose-fermi mixture
subjected to a square lattice and found that the instability
corresponding to bose-bose phase separation always happens at a
higher temperature than the one corresponding to bose-fermi phase
separation. Moreover, we find both the transition temperature
$T_{\rm \scriptscriptstyle DW}$ of supersolid and the coherence
peak at $k_{\rm \scriptscriptstyle DW}$ are enhanced in the
mixtures studied. These will make the observation of supersolid in
experiments more reachable.

\section*{Acknowledgement}

This work is supported by NSFC Grant No.11275180.

\end{document}